\documentclass[apj]{emulateapj}
\begin{document}
\title{The 8~\micron~Phase Variation of the Hot Saturn HD~149026b}
\author{Heather A. Knutson\altaffilmark{1,2}, David Charbonneau\altaffilmark{1}, Nicolas B. Cowan\altaffilmark{3}, Jonathan J. Fortney\altaffilmark{4}, Adam P. Showman\altaffilmark{5}, Eric Agol\altaffilmark{3}, and Gregory W. Henry\altaffilmark{6}}
\altaffiltext{1}{Harvard-Smithsonian Center for Astrophysics, 60 Garden St., Cambridge, MA 02138}
\altaffiltext{2}{hknutson@cfa.harvard.edu}
\altaffiltext{3}{Department of Astronomy, Box 351580, University of Washington, Seattle, WA 98195}
\altaffiltext{4}{Department of Astronomy and Astrophysics, UCO/Lick Observatory, University of California, Santa Cruz, CA 95064}
\altaffiltext{5}{Lunar and Planetary Laboratory and Department of Planetary Sciences, University of Arizona, Tucson, AZ 85721}
\altaffiltext{6}{Center of Excellence in Information Systems, Tennessee State University, 3500 John A. Merritt Blvd., Box 9501, Nashville, TN 37209}

\begin{abstract}

We monitor the star HD~149026 and its Saturn-mass planet at 8.0~\micron~over slightly more than half an orbit using the Infrared Array Camera (IRAC) on the \emph{Spitzer Space Telescope}.  We find an increase of $0.0227\% \pm0.0066\%$ ($3.4\sigma$ significance) in the combined planet-star flux during this interval. The minimum flux from the planet is $45\% \pm19\%$ of the maximum planet flux, corresponding to a difference in brightness temperature of $480\pm140$~K between the two hemispheres.  We derive a new secondary eclipse depth of $0.0411\% \pm0.0076\%$ in this band, corresponding to a dayside brightness temperature of $1440\pm150$~K.  Our new secondary eclipse depth is half that of a previous measurement ($3.0\sigma$~difference) in this same bandpass by \citet{harr07}.  We re-fit the \citet{harr07} data and obtain a comparably good fit with a smaller eclipse depth that is consistent with our new value.  In contrast to earlier claims, our new eclipse depth suggests that this planet's dayside emission spectrum is relatively cool, with an 8~\micron~brightness temperature that is less than the maximum planet-wide equilibrium temperature.  We measure the interval between the transit and secondary eclipse and find that that the secondary eclipse occurs $20.9^{+7.2}_{-6.5}$~minutes earlier ($2.9\sigma$) than predicted for a circular orbit, a marginally significant result.  This corresponds to $e \cos{(\omega)}=-0.0079^{+0.0027}_{-0.0025}$ where $e$ is the planet's orbital eccentricity and $\omega$ is the argument of pericenter.
                                                        
\end{abstract}


\section{Introduction}\label{intro}

The planet orbiting HD 149026 is unique among the ranks of transiting extrasolar planets.  It has a mass comparable to that of Saturn \citep{sato05,winn08,carter09} but its small radius and correspondingly high average density \citep{sato05,charb06,winn08,nutz09,carter09} suggest that, unlike Saturn, an incredible $50-90\%$ of this planet's mass must exist in the form of a solid icy or rocky core \citep{sato05,fort06,ikoma06,broeg07,burr07a}.  Together with the Neptune-mass planets GJ~436b \citep{butler04,gillon07b} and HAT-P-11b \citep{bakos09}, this places HD~149026b in a class that is distinct from that of the more massive ``hot Jupiter'' transiting planets with their primarily hydrogen-helium compositions. 

In light of its large rock or ice core, it is possible that HD~149026b also has an atmospheric composition that differs significantly from those of the hot Jupiters.  In the solar system, there is a correlation between atmospheric metallicity and the percentage of planet mass that is core \citep[e.g.][]{lodders03}.  For example, Uranus and Neptune have a C/H ratio of $30-40$ times solar while Jupiter's is roughly three times solar.  If this relation holds true for extrasolar planets, it would suggest that HD~149026b may have a significantly metal-enriched atmosphere.  The HD~149026 primary is also metal-enriched ([Fe/H]$=0.36 \pm0.05$, Sato et al. 2005), thus even if HD~149026b's atmosphere simply reflects the composition of its primordial nebula we would expect it to have a slightly higher heavy metal content than a typical hot Jupiter.  

By observing the decrease in light as the planet passes behind its parent star in an event known as a secondary eclipse, we can characterize the properties of HD~149026b's dayside emission spectrum.  \citet{harr07} (hereafter H07) measured HD~149026b's secondary eclipse at 8~\micron~and found a depth of $0.084\%^{+0.009\%}_{-0.012\%}$, corresponding to a brightness temperature\footnote{Brightness temperature is defined in this case as the temperature required to match the observed planet-star flux ratio in the 8~\micron~\emph{Spitzer} IRAC bandpass assuming that the planet radiates as a blackbody and using a Kurucz atmosphere model for the star.} of $2300\pm200$~K for the planet.  This is significantly higher than expected, as HD~149026b has an equilibrium temperature\footnote{Equilibrium temperature is calculated by assuming that the planet absorbs all incident flux (i.e., zero albedo) and then re-radiates that energy over its entire surface as a blackbody of the stated temperature.} of only $1740$~K (H07).  In order to reproduce this flux in the case where the planet emits as a blackbody, we must assume that the planet absorbs and then instantaneously re-radiates all of the incident flux from its star on the day side.  In this scenario the planet's night side would have effectively zero flux; this follows from the requirement that the energy emitted by the planet does not exceed the energy it absorbs (note that residual heat from formation is not expected to play a significant role in the energy budgets of close-in planets).  This is not unreasonable, as the time scale for tidal locking of close-in planets such as HD~149026b is considerably shorter than the ages of these systems \citep{guillot96,lubow97}.  Uneven irradiation of the upper atmosphere could produce a large observed day-night temperature gradient if radiative time scales are much less than advective time scales at the level of the mid-IR photosphere (i.e., flux is absorbed and re-emitted by gas on the dayside in less time than it takes that same parcel of gas to reach the nightside hemisphere).  If the planet's spectrum differs significantly from that of a blackbody with enhanced emission at 8~\micron, the constraints on the night-side emission becomes correspondingly less stringent.  If we compare HD~149026b's 8~\micron~flux to the predictions from 1D atmosphere models for the planet \citep[e.g.][]{fort06,burr08}, we find that models with a temperature inversion and water features in emission at 8~\micron~still require a hot day side and correspondingly large day-night temperature gradient in order to match the 8~\micron~flux observed by H07.  

In this paper we characterize HD~149026b's 8~\micron~phase variation and corresponding day-night temperature gradient by monitoring the system continuously over slightly more than half an orbit, beginning before the transit and ending after the secondary eclipse.  By measuring the increase in brightness as the planet's dayside rotates into view, we can estimate the day-night temperature gradient and constrain corresponding atmospheric circulation models for the planet.  We have previously published similar observations of HD~189733b at 8 and 24~\micron~\citep{knut07,knut09a} where we invert the observed phase variation to produce a longitudinal temperature profile for the planet.  These observations revealed that HD~189733b has a warm night side (approximately $200$~K cooler than the $1250$~K day side), but lower-cadence observations of the non-transiting planets $\upsilon$~And~b \citep{harr06} and HD~179949b \citep{cow07} suggest that other hot Jupiters may have large day-night temperature gradients.  HD~149026b's lower mass and increased core fraction make it qualitatively different than any of these systems, and there are no published general circulation models for this planet analogous to those available for HD~189733b and HD~209458b \citep[e.g.][]{show09}.  As a result, we have no a priori predictions for the nature of the day-night circulation on HD~149026b other than the constraint provided by the secondary eclipse measurement of H07.

These same data also allow us to search for time-varying properties of the system by comparing the depths and relative times of our transit and secondary eclipse to previously published 8~\micron~\emph{Spitzer} observations of the planet's transit \citep[][obtained in 2007]{nutz09} and secondary eclipse (H07, obtained in 2005).  If the planet's orbit is changing over time, perhaps as the result of interactions with an unknown second planet in the system, it could cause deviations in the timing of successive transits \citep{miralda02,holman05,agol05}.  Comparing our 8~\micron~planet-star radius ratio to the value obtained by \citet{nutz09} allows us to search for changes in the effective radius of the planet at 8~\micron~and, if the two values are consistent, obtain an improved estimate for this quantity.  By measuring the interval between the transit and secondary eclipse we can determine a value for $ecos(\omega)$ where $e$ is the planet's orbital eccentricity and $\omega$ is the argument of pericenter, and compare it to the corresponding value from H07.  Finally, by comparing the 8~\micron~secondary eclipse depth from our observations to that of H07, we can search for variations in the planet's 8~\micron~dayside flux. 

In \S\ref{obs} we describe our treatment of the data, including our initial photometry, the use of a new preflash technique to remove the detector ramp, and our fits to the transit, secondary eclipse, and phase curve data.  In \S\ref{discussion} we discuss the results of these fits and check for variability in the relative depth and timing of our transit and secondary eclipse as compared to previously published data.  We also compare our day- and night-side fluxes to the predictions from 1D atmosphere models and calculate an energy budget for the planet.  In \S\ref{conclusions} we summarize our results and discuss the potential impact of upcoming observations of this planet's secondary eclipse and phase curve at shorter wavelengths.

\section{Observations}\label{obs}

We monitored HD~149026b continuously for 41~hours using 8~\micron~IRAC subarray \citep{faz04} on the \emph{Spitzer Space Telescope} \citep{wern04}, from UT 2008 May $11-12$.  Our observations began 2.5 hours before the start of the transit and finished 44 minutes after the end of the secondary eclipse.  We used an integration time of 0.4~s, well below saturation for this star, and obtained a total of 344,704 images.   

\subsection{Initial Photometry}\label{init_phot}

Because HD 149026 is a bright star ($K=6.819$, one-third the brightness of HD~189733 in the infrared) relative to the background at these wavelengths, we calculate the flux from the star in each image using aperture photometry with a radius of 3.5 pixels.  We determine the position of the star in each image as the position-weighted sum of the flux in a circular region with a radius of five pixels (rounded to the nearest pixel) centered on the approximate position of the star.  We calculate position estimates with larger and smaller apertures, but find that a five pixel aperture produced the lowest RMS in the resulting time series.  Similarly, we find that decreasing our photometric apertures to values smaller than 3.5 pixels led to position-dependent flux losses, while the time series from larger apertures become increasingly noisy.  We repeat our analysis for apertures ranging from $3.5-7$ pixels and find that our results for the measured secondary eclipse depth and phase variation amplitude are consistent in all cases.  We also try fitting a point spread function (psf) to the images, using either the in-flight point response functions generated from calibration test data \footnote{Available at http://ssc.spitzer.caltech.edu/irac/psf.html} or a psf derived from our own data, and conclude that aperture photometry produces the optimal results.

Because the subarray is only $32\times32$ pixels, care is needed in the choice of regions for background flux estimates.  We exclude all pixels within a circular region with a radius of 9 pixels centered on the position of the star, as well as the top row of the array which has consistently low flux values, and trim $3\sigma$ outliers from the remaining set of pixels.  We then make a histogram of the fluxes in the remaining pixels and fit a Gaussian function to the central 7 of 10 bins in this distribution (further reducing the effects of any remaining high- or low-flux outliers in the distribution) to estimate the mean background level in each individual image.  We obtain identical results if we include the top row of pixels, increase the size of the exclusion region in the center of the array, or exclude an additional four rows from $y=11-16$~corresponding to particularly bright diffraction spikes on the star's psf.  We find that the background flux during our observations is consistently below zero, with a median value of $-0.053$~MJy~Sr$^{-1}$, suggesting that the sky dark used for these observations is over-subtracting the background, but this has no effect on our final photometry provided that the negative background value is correctly subtracted from each image before estimating the flux from HD~149026.

We calculate the JD value for each image as the time at mid-exposure, and apply a correction to convert these JD values to the appropriate BJD, taking into account Spitzer's orbital position at each point during the observations.  The background and stellar fluxes in the first 10 images and the 58th image in each set of 64 have consistently lower values than the rest of the exposures, but we find that this behavior disappears after the background is subtracted.  As a check we exclude these low frames and repeat our analysis, and find that we obtain consistent results (secondary eclipse depth and phase variation amplitude) in both cases.  We include these frames in our final analysis.

\begin{figure}
\epsscale{1.2}
\plotone{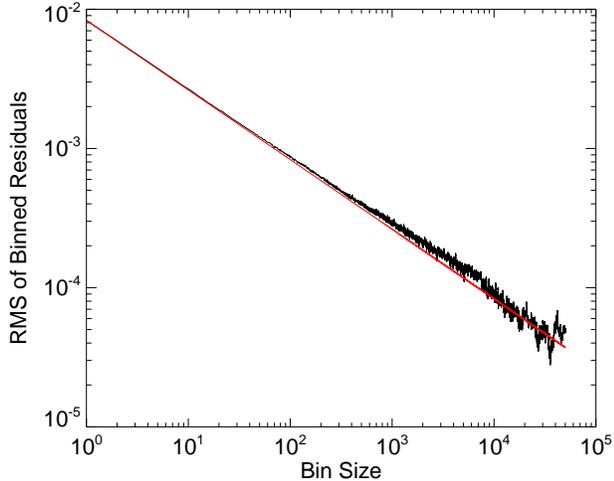} %final_rms_plot.ps
\caption{This figure shows the RMS of the time series between the end of the transit and the start of the secondary eclipse ($10<dt<35$~hours) over a range of bin sizes (black line, bins up to 50,000 points).  We have fitted and removed a linear function of time from these data to correct for the effects of the planet's phase variation before binning the data and calculating the resulting RMS.  The solid red line shows the RMS of the unbinned data scaled by $\sqrt{n}$ where $n$ is the number of points in each bin.  The increased noise for bin sizes between $1000-4000$ points ($10-30$ minute time scales) may be related to the pointing jitter of the telescope, which shifts back and forth by approximately 0.1 pixels in $x$ and $y$ over 1 hour cycles.  Although we find no evidence for a direct relation between the measured $x$/$y$ positions of the star on the array and the corresponding stellar fluxes, the fact that this excess noise disappears for bin sizes larger than 8,000 points (1 hour intervals) suggests these two phenomena may be related. \label{rms_plot}}
\end{figure}

Our final time series has a number of high-flux outliers produced by the presence of transient hot pixels in the array.  We remove these points by discarding images where the $x$ position, $y$ position, or total flux from the star in a 3.5-pixel aperture differed by more than $3\sigma$ from the median value for these quantities in each set of 64 images (because we calculate the star's position as the position-weighted sum of the flux in a given region, transient hot pixels may affect these estimates even if they are not included in the final photometry).  We trim $0.78\%$ of the images using this method, and see no evidence for remaining outliers in the final time series.  We calculate the RMS of the points in each bins in the phase curve plotted in Fig. \ref{phase_curve_plot} and find that the average RMS in these bins is $0.826\%$, which is a factor of 1.27 higher than the predicted photon noise from the star alone (we do not include the noise contribution from the background flux because this flux is negative in our images, and we do not have a reliable estimate of its true value with a correct sky dark subtraction).  This is consistent with previous \emph{Spitzer} photometry in this bandpass for GJ~436, HD~149026, and HD~189733 \citep{dem07,harr07,knut07}, all of which have roughly comparable fluxes in this bandpass.  We find that this noise is reasonably white; Fig. \ref{rms_plot} shows the RMS of the time series between the end of the transit and the start of the secondary eclipse ($10<dt<35$~hours) over a range of bin sizes up to 50,000 points after the data has fitted with a linear function of time which was then removed to ensure that the real astrophysical change in the planet's flux does not inflate the resulting RMS estimates.

\begin{figure}
\epsscale{1.2}
\plotone{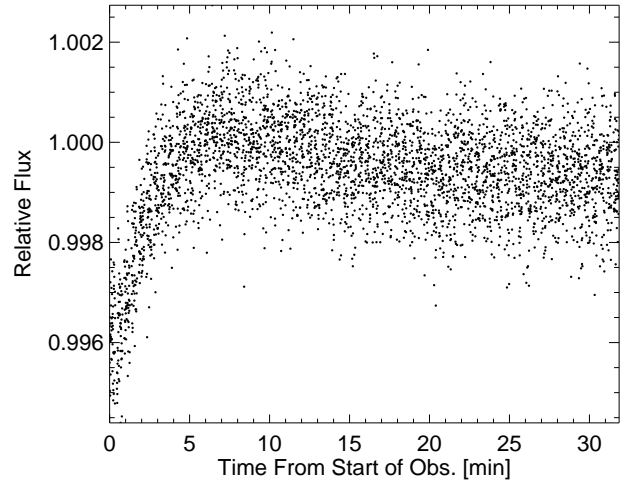} %preflash.ps
\caption{Time series for a 3.5 pixel region centered on the approximate position of HD~149026 (corresponding to the center of the subarray) over the 30 minutes of our preflash observations of M17.  The average flux in this region is 7400~MJy Sr$^{-1}$, but we see an identical behavior  for regions near the corners of the array with average fluxes ranging from $3900-8200$~MJy Sr$^{-1}$. This suggests that we are effectively saturating out the detector ramp in the first five minutes of our preflash observations.\label{preflash}}
\end{figure}

\subsection{Preflashing and the Detector Ramp Effect}\label{preflash_discussion}

After determining the optimal method for estimating the fluxes in each bandpass, we must remove any remaining detector effects from the data.  There is a well-documented detector effect \citep{harr07,knut07,knut08,charb08} in this array that causes the effective gain (and thus the measured flux) in individual pixels to increase over time.  This effect has been referred to as the ``detector ramp'', and has also been observed to occur in the IRS 16~\micron~peak-up and MIPS 24~\micron~arrays, which are made from the same material \citep{dem06,charb08}.  The IRAC 5.8~\micron~array also shows a related behavior, with an initial upward asymptote in the measured flux over the first $30-60$ minutes of observation followed by a more gradual downward slope \citep{charb08,knut08,knut09b,mach08}.  The size of this effect depends on the illumination level of the individual pixel.  Pixels with high illuminations ($>$250 MJy Sr$^{-1}$ in the 8~\micron~channel) will converge to a constant value within the first hour of observations, whereas lower-illumination pixels will show a linear increase in the measured flux over time with a slope that varies inversely with the logarithm of the illumination level.  It has been suggested that this effect may be the result of charge-trapping in the array, with higher-illumination pixels filling their wells more quickly than lower-illumination pixels.  

We mitigated this effect by observing M17, a nearby open cluster with diffuse bright emission at 8~\micron, centered on the position $\alpha=18^h 20^m 28.0^s$, $\delta=-16\degr 12\arcmin 19.5\arcsec$ for 32~minutes immediately prior to the start of our science observations.  This preflash is designed to saturate out the ramp effect by exposing the array to a uniformly bright source with fluxes that are approximately $10\times$ higher than the peak flux in the center of HD~149026's psf.  The region of M17 that we use for our preflash has fluxes between $3500-8000$~MJy Sr$^{-1}$ in the IRAC 8~\micron~subarray aperture, with an average flux of 7400~MJy Sr$^{-1}$ in a circular region with a radius of 3.5 pixels centered on the position of HD~149026 in our science images. We use the same exposure time and subarray observing mode for our preflash that we use for our observations of HD~149026, obtaining a total of 4480 preflash images.  The peak flux in these images occurs at the position of a star in the corner of the array with a maximum flux of 13,000~MJy Sr$^{-1}$ or 100,000 e$^{-}$, corresponding to half of the full well depth in this array.  Because we are well below saturation, we do not expect any latency effects in our subsequent science images.

We evaluate the effectiveness of our preflash by calculating the total flux in an aperture with a 3.5 pixel radius centered on (1) the center of the array in the preflash images, corresponding to the position of HD~149026 in our science observations, (2) a region near the corner of the array in the preflash images with a lower flux around 3500~MJy Sr$^{-1}$, and (3) a bright star in the opposite corner of the array in the preflash images with a peak flux of 13,000~MJy Sr$^{-1}$.  We find that in all cases there is a linear increase in the measured flux during the first five minutes of observations with a total amplitude of 0.4\% (see Figure \ref{preflash}), followed by a slight decrease in the measured flux and then remaining effectively constant for the final 15 minutes of observations.  Based on these data we conclude that the effectiveness of our preflash is independent of the brightness of our preflash source above a minimum threshold value of $4000$~MJy Sr$^{-1}$, and that the charge traps across our entire $32\times32$ pixel subarray are filled in the first five minutes of our preflash observations.
  
\begin{figure}
\epsscale{1.2}
\plotone{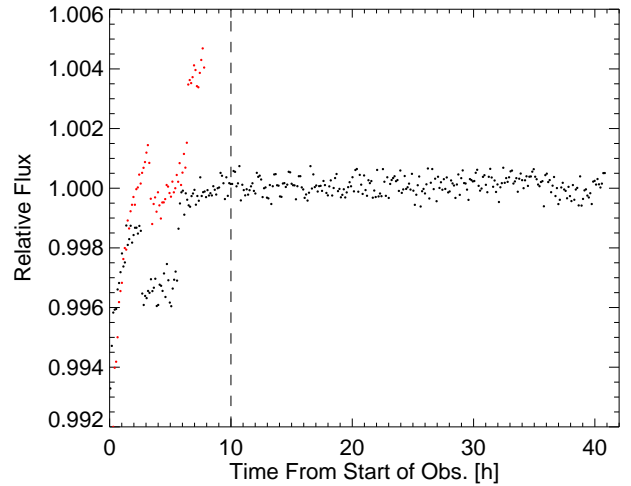} %raw_phot.ps
\caption{Final time series for HD 149026 (black points) over the 41 hours of our observations, with data binned in 428~s intervals containing approximately 1000 points per bin.  There is still a small ramp at early times, but its amplitude is smaller than the equivalent ramp with no preflash and the ramp converges to a constant value within the first 10 hours of observations.  For comparison we show equivalently binned photometry for a transit observation of HD 149026b observed on UT 2007 August 14 in this same bandpass with identical 0.4~s exposures in the 8~\micron~IRAC array \citep{nutz09}.  Unlike our data these older data were not preflashed, and the increased slope of the resulting photometry indicates the amplitude of the unmitigated detector ramp for this star.\label{raw_phot}}
\end{figure}

When we plot our final time series for HD~149026 (Fig. \ref{raw_phot}), we find that the preflash has moderated the ramp effect in our science data but it has not removed it entirely.  A comparison to identical observations of HD~149026 \citep[8~\micron~IRAC array, 0.4~s integration times,][]{nutz09} from 2007 reveals that the amplitude of the ramp in our preflashed data is considerably smaller than the equivalent ramp in these older data.  We speculate that the smaller ramp in our data may be the result of a slow leakage of charge from the filled charge traps during the 14.4 minute interval between the end of our preflash and the start of our science data that partially resets the detector ramp effect.  We note that  unlike the original unmitigated ramp, we see no variation in the slope of this new ramp with increasing aperture, and the background fluxes show a slight \emph{decrease} in flux during the first few hours of observations.

Our data do not show any evidence for a detector ramp after the first 10 hours of observations (see \S\ref{phase_var_fits} for a more detailed discussion on this point).  We account for the ramp at early times ($<10$ hours after the start of observations) by fitting the following quadratic function of $\ln{(dt)}$ simultaneously with our transit fit:

\begin{equation}\label{eq1}
f=c_1+c_2\ln{(dt+c_4)}+c_3\left(\ln{(dt+c_4)}\right)^2
\end{equation}

where $f$ is the measured flux, $dt$ is the elapsed time in days since the start of the observations, and $c_1-c_4$ are free parameters in the fit.  We also tried fitting this slope with a linear function of $\ln{(dt)}$ or a simple exponential function, but found that Eq. \ref{eq1} provided the best fit to the data. 

\begin{figure}
\epsscale{1.2}
\plotone{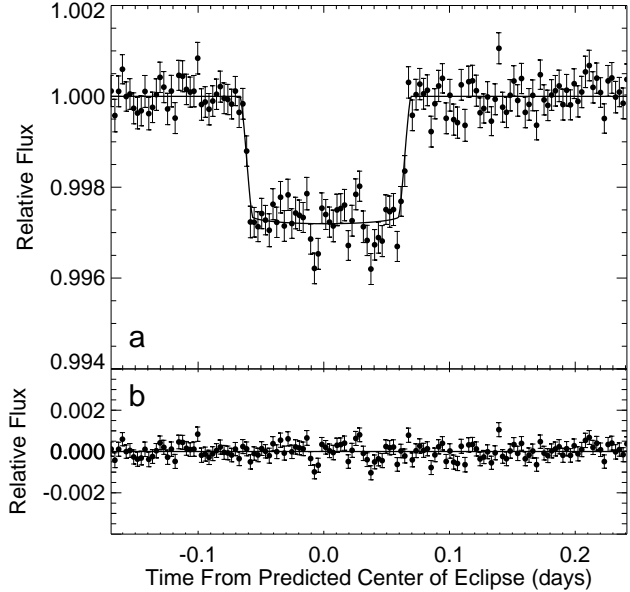} %transit_plot_final.ps
\caption{This plot shows the first ten hours of data with the detector ramp removed and the best-fit transit light curve overplotted In this plot we use the fit with $a/R_*$ and $i$ fixed to their best-fit values from \citet{carter09}.  The data are binned in 259~s intervals containing approximately 600 points per bin.  The uncertainties on each individual bin are calculated as the standard deviation of the flux values in that bin divided by the square root of the number of points.  Data during the center of transit show several above-average deviations from the best-fit transit light curve; the time scale of these deviations is similar to the planet crossing time scale, and such deviations could be caused by the presence of star spots intersecting the planet's path across the surface of the star.\label{transit_plot}}
\end{figure}

\begin{figure}
\epsscale{1.2}
\plotone{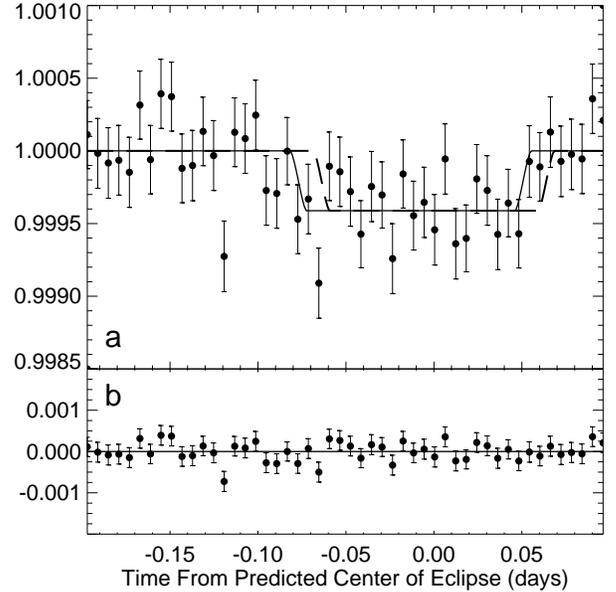} %sec_eclipse_plot_final.ps
\caption{This plot shows the 7.2 hours of data used for the secondary eclipse fit, beginning 0.2 days from the predicted center of the eclipse).  The data are binned in 517~s intervals containing approximately 1200 points per bin.  The uncertainties on each individual bin are calculated as the standard deviation of the flux values in that bin divided by the square root of the number of points.  The two overplotted curves show the best-fit eclipse light curve with best-fit transit time (solid line) or with the time fixed to its predicted value (dashed line).  We find that the secondary eclipse occurs $20.9^{+7.2}_{-6.5}$ minutes earlier than predicted for a circular orbit (see discussion in \S\ref{eclipse_timing}).\label{sec_eclipse_plot}}
\end{figure}

\subsection{Eclipse Fits}\label{eclipse_fits}

We fit both Eq. \ref{eq1} and the transit curve to the first ten hours of data simultaneously using a Markov Chain Monte Carlo (MCMC) method \citep[e.g.][]{tegmark04,ford05,winn07,burke07}.  We exclude the first 30 minutes of data from our fit, as the ramp is particularly steep during this time, and set the uncertainties on individual points equal to the RMS variance of the time series between $0.8-1.0$ days after the start of the observations (this is the middle of the time series, well after the transit but before the start of the secondary eclipse, and we have selected a shorter segment to avoid inflated errors due to the planet's phase variation).  We fix the four parameter nonlinear limb-darkening coefficients \citep{claret00} to the values used by \citet{nutz09}, which are derived by fitting the appropriate limb-darkening law to a Kurucz stellar atmosphere model for a $T_{eff}=6250$~K, $log(g)=4.5$,[Fe/H]$=0.3$ star \citep{kurucz79,kurucz94}.  We calculate our transit light curve using the routines from \citet{mand02} and fit for four free parameters in addition to the four parameters in Eq. \ref{eq1}, including: the planet-star radius ratio $R_P/R_*$, the ratio of the planet's orbital distance to the stellar radius $a/R_*$, the orbital inclination $i$, and the transit time.  These four parameters are independent of assumptions about the properties of the star, and can be derived directly from the light curve.  Using Kepler's law together with the planet's orbital period and the stellar mass, we can then derive the corresponding values for the stellar and planetary radii $R_*$ and $R_P$.  This gives us a total of eight free parameters in our fit to the transit, and we run our Markov Chain for a total of $10^6$ steps.  

After running the chain, we search for the point in the chain where the $\chi^2$ value first falls below the median of all the $\chi^2$ values in the chain (i.e. where the code had first found the best-fit solution), and discard all the steps up to that point.  We take the median of the remaining distribution as our best-fit parameter, with errors calculated as the symmetric range about the median containing 68\% of the points in the distribution.  We find that $a/R_*$ and $i$ are highly correlated and have asymmetric histograms, therefore for these two parameters we calculate separate positive and negative error bars.

\begin{deluxetable*}{lrrrrcrrrrr}
\tabletypesize{\scriptsize}
\tablecaption{Best-Fit Transit and Secondary Eclipse Parameters \label{eclipse_param}}
\tablewidth{0pt}
\tablehead{
\colhead{Parameter} & \colhead{Independent Fit\tablenotemark{a}}  & \colhead{\citet{carter09} Fit\tablenotemark{b}}}
\startdata
Transit Parameters & & \\
$R_P/R_*$ & $0.05216\pm0.00078$ & $0.05253\pm0.00076$ \\
$i$ ($\degr$) & $88.0^{+1.4}_{-1.9}$ & $84.50$ \\
$a/R_*$\tablenotemark{c} & $7.25^{+0.02}_{-0.70}$ & $5.99$ \\
Center of Transit & $2454597.70721\pm0.00040$~BJD & $2454597.70709\pm0.00039$~BJD \\
O-C (minutes)\tablenotemark{d} & $1.10\pm0.63$ & $0.92\pm0.62$\\
 & & & \\
Secondary Eclipse Parameters & & \\
Depth  & $0.0411\% \pm0.0076\%$ & \\
Center of Eclipse & $2454599.1304\pm0.0048$~BJD & \\
O-C (minutes)\tablenotemark{d,e} & $-20.9^{+7.2}_{-6.5}$ & \\
\enddata
\tablenotetext{a}{These values are from an independent fit to our data alone, with no prior assumptions about $a/R_*$ or $i$.}
\tablenotetext{b}{Here we set fix $a/R_*$ and $i$ to their best-fit values from \citet{carter09}.}
\tablenotetext{c}{The probability distribution for $a/R_*$ is highly asymmetric, so we use the mode instead of the median as our best-fit value and set our uncertainties to the 16th and 84th percentiles of the distribution.}
\tablenotetext{d}{Calculated using the new ephemeris derived in this paper; the uncertainties include the uncertainties in both the observed and calculated transit times.}
\tablenotetext{e}{The O-C for the secondary eclipse is calculated assuming that the secondary eclipse occurs half an orbit after the transit plus an additional 45~s delay from the increased light travel time in the system \citep{loeb05,harr07}.}
\end{deluxetable*}

As noted above, our transit data are virtually identical to that of \citet{nutz09} with the sole exception of a flatter detector ramp (see Fig. \ref{raw_phot}).  \citet{carter09} recently presented a second, independent analysis of the \citet{nutz09} data in combination with new observations of the transit between $1.1-2.0$~\micron~using NICMOS on the \emph{Hubble Space Telescope}.  The authors in this study derive improved constraints on $a/R_*$ and $i$ relative to an analysis of the \emph{Spitzer} data alone.  In order to facilitate comparisons between these two data sets, we repeat our fits with these two parameters fixed to their best-fit values from \citet{carter09} and allow the remaining six parameters (four from Eq. \ref{eq1}, $R_P/R_*$, and the transit time) to vary freely.  We give the results of both fits in Table \ref{eclipse_param} below.

We perform our fit to the secondary eclipse using similar methods; there is no detector ramp in this region of the time series so we limit our out-of-transit normalization to a simple linear function of time which we fit simultaneously with the eclipse.  We select the final 7.2 hours of data used for the secondary eclipse fit (beginning 0.2 days before the predicted center of the eclipse in order to provide an adequate out-of-eclipse baseline).  We find the slope of the linear term in this fit is consistent with zero, and obtain similar results if we fit a constant value to the out-of-eclipse data \citep[see discussion in][]{knut09a}.  We allow both the depth and timing of the secondary eclipse to vary independently, and take the other parameters for the system (planetary and stellar radii, orbital period, etc.) from \citet{nutz09}.  Given the relatively low signal-to-noise of this eclipse as compared to the transit, we conclude that updating the system parameters to the values from \citet{carter09} would have no effect on our final best-fit transit depth and corresponding uncertainties.  As before, we calculate our eclipse curve using the equations from \citet{mand02} with limb-darkening set to zero.  

Figures \ref{transit_plot} and \ref{sec_eclipse_plot} show the final binned data from our fits to the transit and secondary eclipse with the out-of-transit trends removed and the best-fit eclipse curves overplotted.  The corresponding best-fit parameters are given in Table \ref{eclipse_param}.

\subsubsection{Fits to 8~\micron~Secondary Eclipse Data from 2005}\label{harr_data}

As part of this paper we would like to compare our new 8~\micron~secondary eclipse depth to the 8~\micron~secondary eclipse depth from 2005 presented in H07.  Over the past two years there have been a number of changes both to the standard \emph{Spitzer} pipeline and also in the accepted best-practice methods for dealing with these data.  In light of these changes, we decided to evaluate the robustness of the measured 2005 eclipse depth by downloading the images\footnote{These data are publicly available through the \emph{Spitzer} archives.} from 2005 and performing our own independent analysis of these data.  The images we downloaded from the \emph{Spitzer} archives, which are periodically reprocessed with updated versions of the \emph{Spitzer} pipeline, were last updated in June 2006 (version S14.0.0).  We use these images to create an updated photometric time series, following the same methods as those described in \S\ref{init_phot}, and compare the best-fit eclipse depths obtained using these data with those obtained using the photometry from the original H07 paper\footnote{This photometry was provided as part of the H07 Nature paper's supplemental information.}.  We find that we obtain the same eclipse depths using both our new photometry and the original H07 photometry.  As described above, we calculate our aperture photometry using a circle with a radius of 3.5 pixels, but we obtain consistent results using the slightly larger 4.0 pixel aperture selected by H07.

Unlike our data, the H07 data cycles between nine different nod positions, with a total of twelve cycles through these nod positions (see the supplemental methods section of H07 for more information).  This creates a rather complicated time series, as there is both an overall ramp in the data comparable to the one described in \S\ref{preflash_discussion}, and a series of smaller ramps that begin each time the telescope moves to a new pointing position.  We follow the general method described in H07 and fit these effects with a combination of functions, including a function to describe the overall ramp:

\begin{equation}\label{ramp_eq}
f_{ramp}=c_1 \left(1-e^{c_2(dt_{obs}+0.75)}\right)
\end{equation}

where $c_1$ and $c_2$ are free parameters in the fit, $dt_{obs}$ is the elapsed time from the start of the observations in days, and $0.75$ is an arbitrary time offset.  We tried fits in which this time offset was a free parameter, but the additional degree of freedom did not improve the quality of the resulting fit and our Markov chains indicated that the time offset and $c_2$ were highly degenerate.  We selected the value of $0.75$ from the center of the distribution of values where this parameter was allowed to vary freely.  We also ran fits using a quadratic functions of $\ln{dt}$, similar to that given in Eq. \ref{eq1} but with the phase shift $c_2$ in that equation fixed to a value of 0.02 days, and found that the exponential function was a better fit to the data (H07 reached a similar conclusion).

As mentioned above, there is a series of smaller ramps in these data that reset each time the telescope moves to a new pointing position.  We fit these ramps with a function of $\ln{dt}$, comparable to that of Eq. \ref{eq1}:

\begin{equation}\label{nod_eq}
f_{nod}=1+c_3\ln{(dt_{nod}+c_4)}
\end{equation}

where $c_3$ and $c_4$ are free parameters and $dt_{nod}$ is the elapsed time in days since the start of observations at the current nod position (this parameter resets to zero each time the telescope repoints to the next nod position in the cycle).  Our choice to fit this ramp with an analytic function differs from that of H07, who defined the ramp at each nod position using spline interpolation through the photometry from twelve images in each visit where the final three observations were fixed to unity (see \S5 in the supplemental information section in H07).

We also include constant flux offsets $c_5-c_{12}$ at eight of the nine nod positions (the offset at the first nod position is fixed to zero) as free parameters in our fits in order to account for position-dependent differences in the measured fluxes across the array.  H07 do the same in their fits.  We allow both the depth and timing of the secondary eclipse to vary independently, and take the other parameters for the system (planetary and stellar radii, orbital period, etc.) from \citet{nutz09} as described in \S\ref{eclipse_fits}.  This choice differs from that of H07, who describe their eclipse with four free parameters (eclipse duration between half-light points, center time, depth, and limb-crossing duration), as the uncertainties in the eclipse duration and length of ingress/egress were much higher at the time this paper was written.

We carry out our fits using a Markov Chain Monte Carlo method, identical to the one described in \S\ref{eclipse_fits} with a total of 14 free parameters (H07  have 23 free parameters in their fit):

\begin{eqnarray}
f_{measured}&=&f_{ramp}(c_1,c_2)*f_{nod}(c_3,c_4) \\
& &*f_{eclipse}(c_{13},c_{14})+f_{offset}(c_5-c_{12}) \nonumber
\end{eqnarray}

where $f_{measured}$ is the measured flux, $f_{ramp}$ is given by Eq. \ref{ramp_eq} and describes the overall ramp in the data, $f_{nod}$ is given by Eq. \ref{nod_eq} and describes the smaller ramp at each new nod position, $f_{offset}$ is the constant flux offset for the last 8 of 9 total nod positions, and $f_{eclipse}$ is the secondary eclipse function (calculated the same way as the secondary eclipse in \S\ref{eclipse_fits}) where the eclipse depth $c_{13}$ and the time of center of eclipse $c_{14}$ are left as free parameters.  We set the uncertainty on each individual point equal to the same value used for our new photometry in \S\ref{init_phot} (25.790 MJy Sr$^{-1}$ for a 3.5 pixel aperture or 26.804 MJy Sr$^{-1}$ for a 4.0 pixel aperture), as this reflects the actual RMS achieved for this star in the absence of any significant detector effects.  Our MCMC fit is different than the fitting method employed by H07, who used the Levenberg-Marquardt method to minimize their $\chi^2$ and determine the corresponding best-fit solution, with a bootstrap method with 17,030 trials (as opposed to our $10^6$ step Markov chain) to estimate the corresponding uncertainties.

\begin{figure}
\epsscale{1.2}
\plotone{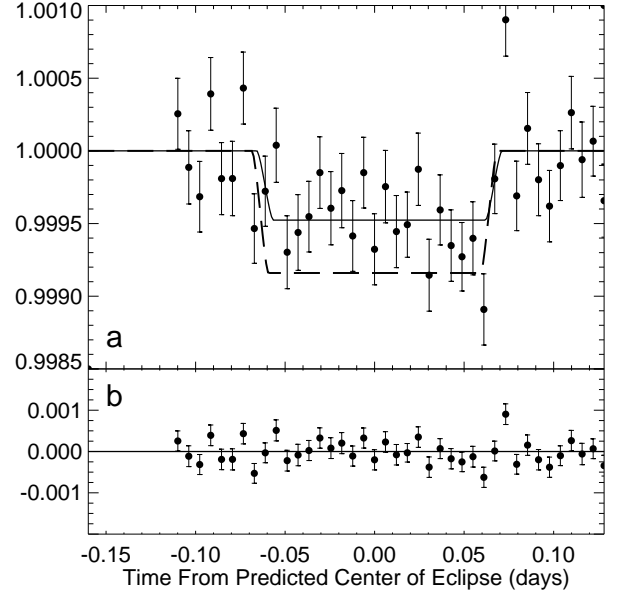} %.ps
\caption{This plot shows the updated photometry for an 8~\micron~eclipse of HD~149026b observed on UT 2005 August 24, originally published by \citet{harr07}.  The time series shown in this plot is calculated our own photometry pipeline with a 4.0 pixel radius aperture and has detector effects removed.  These observations lasted 6.0 hours, but we have plotted the same time interval (7.2 hours) as in Fig. \ref{sec_eclipse_plot} to facilitate comparisons.  The data are binned in 527~s intervals containing approximately 1200 points per bin.  The uncertainties on each individual bin are calculated in the same way as Fig. \ref{sec_eclipse_plot}.  The two overplotted curves show the best-fit eclipse light curve from our analysis (solid line) and an eclipse curve calculated using the best-fit eclipse depth and time from H07 (dashed line).  We conclude that these data are consistent with the secondary eclipse depth from our 2008 data, and find no evidence for time variability in the planet's 8~\micron~flux.\label{harr_plot}}
\end{figure}

\begin{figure*}
\epsscale{0.8}
\plotone{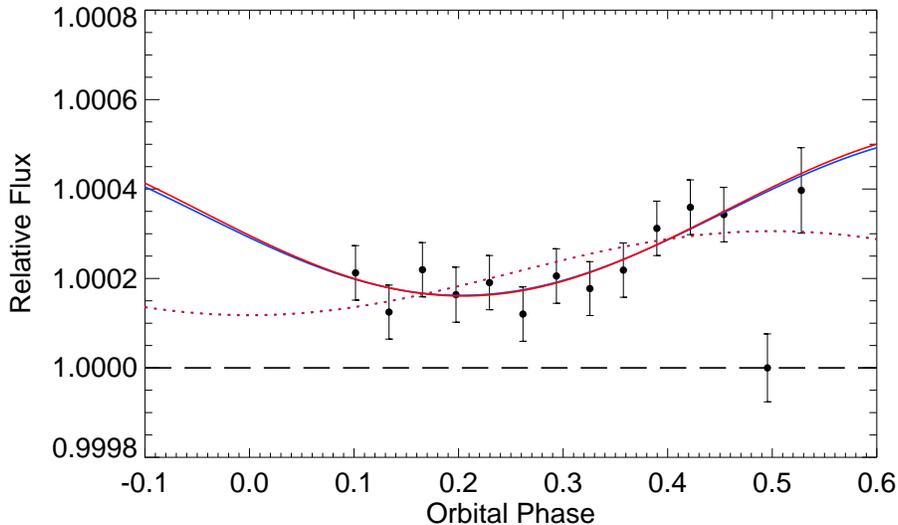} %final_phase_curve_plot.ps
\caption{This plot shows the data used in the phase curve fit, beginning 10 hours after the start of the observations (to avoid the initial detector ramp) and excluding data within 8 minutes ($2\sigma$ uncertainty in the original best-fit eclipse time, see \S\ref{eclipse_timing}) of the start and end of the best-fit eclipse solution.  The stellar flux as measured at the center of the secondary eclipse has been normalized to unity (dashed line).  The data before the start of the secondary eclipse are binned in 2.2 hour intervals containing approximately 20,000 points per bin, and the error bars for each bin are calculated as the standard deviation of the flux values in that bin divided by the square root of the number of points.  This is a slight over-estimate of the uncertainties on individual points, as the decreasing RMS for bin sizes greater than 1 hour (Fig. \ref{rms_plot}) indicates that there is a component of correlated noise, perhaps connected to the pointing jitter of the telescope, that averages out on longer time scales.  Data after the end of the secondary eclipse are binned into a single point, and we show the best-fit secondary eclipse depth with $1\sigma$ uncertainties at the center of the best-fit eclipse solution for comparison.  The solid red line is the best-fit two-hemisphere model where the locations of the two hemispheres are allowed to vary freely in the fit, and the dotted red line is the same model with the locations of the two hemispheres fixed on the substellar and antistellar points.  The additional degree of freedom in the first (solid line) model reduces the $\chi^2$ of the fit by $6.2$, a clear improvement.  The two blue lines are for model fits in which the day-night flux distribution varied as the cosine of the longitude from the substellar point with (solid line) and without (dotted line, located underneath the red dotted curve) an additional phase shift as a free parameter in the fit.  The two-hemisphere and sinusoidal models are indistinguishable at the precision of the data.\label{phase_curve_plot}}
\end{figure*}

As discussed in \S\ref{init_phot}, images taken in subarray mode are stored in sets of 64 and the background and stellar fluxes in the first 10 images and the 58th image in each set of 64 have consistently lower values than the rest of the exposures.  For our new data we found that this effect was removed by the background subtraction, but as a check we repeat our fits to the updated photometry for the 2005 data with and without these images.  We find that we obtain consistent eclipse depths when we trim the 58th image alone (an eclipse depth of $0.055\% \pm0.007\%$), trim the 1st five images and 58th image (an eclipse depth of $0.057\% \pm0.007\%$), or the 1st 10 images and the 58th image (an eclipse depth of $0.060\% \pm0.008\%$).  If we change our method for removing transient hot pixels from the time series from the one described at the end of \S\ref{init_phot} to a simple flux cutoff (any images with fluxes lower than 3000 MJy Sr$^{-1}$ or higher than 3230 MJy Sr$^{-1}$ for a 3.5 pixel radius aperture are removed, where the flux cutoffs are chosen using a simple visual inspection of the time series), we obtain an even deeper eclipse depth of $0.086\%^{+0.016\%}_{-0.020\%}$.  Using a 4.0 pixel aperture and our original bad pixel filtering method, and either trimming both the 58th image and 1st 10 images in each set of 64 or leaving those images in, we obtain eclipse depths of $0.052\% \pm0.009\%$ and $0.048\% \pm0.008\%$, respectively.  If we allow the time offset of $0.75$ in Eq. \ref{ramp_eq} to vary freely in our fits with a 3.5 pixel aperture and trim the 1st 10 images and the 58th images we obtain a slightly deeper eclipse depth with larger uncertainties ($0.076\%^{+0.014\%}_{-0.020\%}$).  However, this fit is not stable, as we obtain a depth of $0.058\%^{+0.009\%}_{-0.012\%}$ if we include the 1st 10 frames in this fit, and if we use photometry with a slightly larger 4.0 pixel aperture instead of 3.5 pixels we obtain a value of $0.052\%\pm0.008\%$ for the eclipse depth.  We plot one example of our binned photometry (best-fit eclipse depth of $0.048\% \pm0.008\%$ with detector effects removed and best-fit eclipse curve overplotted in Fig. \ref{harr_plot}.  The fit plotted in this figure uses photometry with a 4.0 pixel aperture, our best bad pixel filtering method, includes the first 10 and 58th images in our fit, and fixes the time offset in Eq. \ref{ramp_eq} to 0.75.

The diversity of eclipse depths ($0.05\%-0.09\%$) obtained in these fits suggest that the final result is sensitive to our specific choice of functions, fitting routines, and bad pixel trimming methods.  This is not surprising in light of the many degrees of freedom required to fit the detector effects in these data.  As noted above, we obtained consistent results using both our updated version of the 2005 photometry (with our own photometry routines) and the original photometry provided by H07.  

As a final test we compare the $\chi^2$ value for the same kind of fit plotted in Fig. \ref{harr_plot}, but using the original H07 photometry instead of our new photometry, to the $\chi^2$ value of the best-fit solution obtained by H07.  We trim the same images from our final time series as H07; the only difference is that we set the uncertainty on individual points equal to the average uncertainty from H07, rather than using the vector of individual uncertainties provided by H07.  Our best-fit solution as determined from a $10^6$ step Markov chain, has an eclipse depth of $0.050\% \pm0.008\%$ and a reduced $\chi^2$ value of $0.67359847$ with $48,145$ points in the fit and 14 degrees of freedom.  This is slightly smaller than the reduced $\chi^2$ of $0.67368895$ obtained by H07 with the same number of points and 23 degrees of freedom (J. Harrington 2009, private communication), but the difference is not statistically significant. 

Based on this analysis, we conclude that these data are consistent with a shallower eclipse depth than that published by H07.  As we discuss in more detail in \S\ref{depth_comparison}, we find that there is no evidence for variability in the planet's 8~\micron~flux over the three-year interval between these data and our new observations.

\subsection{Phase Curve Fits}\label{phase_var_fits}

Due to the small amplitude of the secondary eclipse depth ($0.041\% \pm0.008\%$), a robust detection of the phase variation of HD~149206b in these data presents a considerable challenge.  For comparison, the measured amplitude of the phase variation for HD~189733b, whose host star is three times brighter than HD~149026 in this bandpass, was $0.12\% \pm0.02\%$ \citep{knut07}. This is five times larger than the best-fit amplitude for HD~149026b's phase variation given below.

\begin{deluxetable*}{lrrrrcrrrrr}
\tabletypesize{\scriptsize}
\tablecaption{Minimum and Maximum Hemisphere-Averaged Fluxes \label{temp_table}}
\tablewidth{0pt}
\tablehead{
\colhead{Parameter} & \colhead{Planet-Star Flux Ratio}  & \colhead{Brightness Temperature\tablenotemark{a}}}
\startdata
Linear Fit to Phase Curve & & \\
($\chi^2=228872.7$, 228920 points \& 2 DOF) & & \\
$F_{max}$\tablenotemark{b} & $0.0411\%\pm0.0076\%$ & $1440\pm150$~K \\
$F_{min}$\tablenotemark{c} & $0.0185\%\pm0.0100\%$ & $960\pm240$~K \\
$F_{min}/F_{max}$ & $45\%\pm19\%$ &  \\
$T_{max}-T_{min}$ &  & $480\pm140$~K \\
 & & & \\
Two Hemisphere Model Fit & & \\
($\chi^2=228867.6$, 228920 points \& 3 DOF) & & \\
$F_{max}$\tablenotemark{d,e} & $0.0539\%\pm0.0138\%$ & $1680\pm260$~K \\
$F_{min}$\tablenotemark{e} & $0.0161\%\pm0.0080\%$ & $910\pm200$~K \\
$F_{min}/F_{max}$ & $30\%\pm13\%$ &  \\
$T_{max}-T_{min}$ &  & $780\pm240$~K \\
\enddata
\tablenotetext{a}{Here we assume the planet emits as a blackbody and solve for the temperature that reproduces the observed planet-star flux ratio in our 8~\micron~bandpass.  We use a PHOENIX NextGen model atmosphere \citep[e.g.][]{haus99} for the star with an effective temperature of 6160 K and log(g)$=4.19$; this is the same model used to calculate the planet-star flux ratios for the 1D models in Figures \ref{spectrum} and \ref{night_spectrum}.}
\tablenotetext{b}{The maximum hemisphere-averaged flux (given as a percentage of the stellar flux) in this fit is set equal to the dayside flux as determine from the secondary eclipse depth.}
\tablenotetext{c}{The uncertainty in the minimum hemisphere-averaged flux includes the uncertainties from both the secondary eclipse depth estimate and the relative increase in flux (in this case, the slope of the line in the phase curve fit).}
\tablenotetext{d}{The maximum flux in this fit occurs after the end of our observations, and the uncertainties in its best-fit value are correspondingly high.}
\tablenotetext{e}{The uncertainty in the minimum and maximum hemisphere-averaged fluxes as determined from the two-hemisphere model fit to the phase curve (\S\ref{phase_var_fits}) includes the uncertainties from both the secondary eclipse depth estimate and the relative increase/decrease in flux from the secondary eclipse to the maximum/minimum hemisphere-averaged flux value; the values quoted in the text of \S\ref{phase_var_fits} do not include the additional uncertainty from the secondary eclipse depth estimate.}
\end{deluxetable*}

As discussed in \S\ref{preflash_discussion}, we trim the first 10 hours of data from our phase curve fits, removing the part of the time series affected by the residual detector ramp.  In our previous observations of HD~189733b at 8~\micron~\citep{knut07}, long-term trends in the time series were caused by low-illumination pixels at the edges of our aperture.  These pixels converged to a constant value more slowly than the high-illumination pixels at the center of the star's psf, a behavior that could be observed by comparing the slope of the ramp for smaller and larger apertures (because larger apertures contained more low-illumination pixels, the time series using these apertures will have a significant ramp even at late times).  In our new preflashed observations of HD~149026 the behavior of the residual ramp in the data does not appear to be correlated with the illumination level of individual pixels, as we see identical behaviors for apertures ranging from $3.5-7.0$ pixels once the background flux has been subtracted.  We take this as evidence that the low-flux, large-slope component of the ramp has been effectively removed from the data, and proceed to fit the data starting 10 hours into the observations without any further corrections.

We first fit a simple linear function of time to the data, taking all points between $10<t<36.53$~hours after the start of the observations and excluding all data beginning 8 minutes ($2\sigma$ uncertainty in the original best-fit eclipse time, see \S\ref{eclipse_timing}) before the start of ingress and ending 8 minutes after the end of egress in the best-fit secondary eclipse light curve.  This corresponds to a range in orbital phase of $0.085-0.534$.  We find an increase in flux of $0.0227\% \pm0.0066\%$ over this interval, with a significance of $3.4\sigma$, for our best-fit linear solution, which has a $\chi^2$ value of 228,872.69, two degress of freedom, and 228,920 points in the fit.  Moving to a quadratic function decreases the $\chi^2$ value of the fit by 4.3, indicating that there is detectible curvature in the shape of the observed phase variation.  For the quadratic fit we find a flux minimum located at an orbital phase of $0.20\pm0.17$, corresponding to a central meridian longitude $72\pm61\degr$ west of the antistellar point.

We also fit a more realistic two-hemisphere model \citep{cow08} to these same data.  We first fit a model where we fix the locations of the day and night side hemispheres to be centered on the substellar and antistellar points on the planet, respectively.  Figure \ref{phase_curve_plot} shows the binned data and resulting best-fit phase curve (dotted line).  We find a maximum (dayside) hemisphere-averaged flux of $0.0305\% \pm 0.0031\%$ and a minimum (nightside) hemisphere-averaged flux of $0.0118\% \pm0.0036\%$ from this fit.  Note that this model underestimates the flux at secondary eclipse (this is analogous to the  maximum hemisphere-averaged flux in this case) in order to provide a better fit to the shape of the phase curve prior to the start of the secondary eclipse.  The $\chi^2$ value of this fit is marginally worse than either the linear or quadratic fits ($\delta \chi^2=1.2$ as compared to the linear fit, $\delta \chi^2=5.5$ for the quadratic fit).  We also tried fitting a model in which the day-night flux distribution varied as the cosine of the longitude from the substellar point; this is the simplest version of the sinusoidal class of models described in \citet{cow08}.  The resulting light curve was almost identical to that of the two-hemisphere model, with $\delta \chi^2=0.03$ (a slightly worse fit).

We obtain an improved fit ($\delta \chi^2=-5.1$ as compared to the linear fit and $\delta \chi^2=-0.9$ as compared to the quadratic fit) if we allow the locations of the two hemispheres to vary in our fit (solid line in Fig. \ref{phase_curve_plot}).  In this fit, the maximum and minimum hemisphere-averaged fluxes do not necessarily occur at the points of superior and inferior conjunction, although these two events will always occur exactly half an orbit apart for the case where the planet has a circular orbit.  The maximum and minimum hemisphere-averaged fluxes in this fit are $0.0539\% \pm 0.0115\%$ and $0.0161\% \pm0.0024\%$, respectively.  In this model the centers of the two hemispheres are shifted to central meridian longitudes $73\pm17\degr$ west of the substellar and antistellar points.  The artificially small uncertainty on this phase shift comes from the assumption of a specific physical model (namely two hemispheres of constant surface brightness, producing a phase curve where the maximum flux always occurs $180\degr$ after the minimum flux).  If we relax this restriction by increasing the number of slices in the fit to either 3 ($120\degr$-wide slices) or 4 ($90\degr$-wide slices), we find the faintest slices are located at central meridian longitudes $72\pm49\degr$ and $88\pm40\degr$ west of the antistellar point, respectively.  This suggests that the true uncertainty in the location of the darkest region on the planet should be $40-50\degr$ in longitude, comparable to and slightly smaller than the uncertainty from our original quadratic fit.  Although we obtain consistent values for this location from all of our fits, we conclude that the offset from the antistellar point is not statistically significant.  We also repeat the same fit with a model in which the flux distribution varies as the cosine of the longitude from the substellar point plus a phase offset; as before, the resulting fit was indistinguishable from the best-fit phase-shifted two-hemisphere model ($\delta \chi^2= -0.01$).

In our discussions below we use the linear fit to estimate the increase in flux from the planet beginning 10 hours after the start of our observations and ending after the end of the secondary eclipse and the corresponding ratio of the minimum to the maximum hemisphere-averaged fluxes.  Although the two-hemisphere phase-shifted model provides a better fit to the data, its estimate of the maximum hemisphere-averaged flux involves an extrapolation to times after the end of our observations.  The linear fit makes no such assumptions, and allows us to make a simple statement about the increase in flux observed during the fraction of the planet's orbit spanned by our observations.

\subsubsection{Variability From Starspots}

As discussed in \citet{knut07,knut09a}, changes in the brightness of the planet-hosting star HD~189733 as starspots rotate in and out of view can contribute to the measured changes in flux over the $30-40$ hours of a typical phase curve observation.  Although the amplitude of these variations is reduced at mid-IR wavelengths as compared to visible light, the signal from the planet is small enough that starspot variations cannot be neglected.  HD~189733 is more active than most planet-hosting stars discovered to date, with a rotation period of 13 days and a corresponding $3\%$ flux variation in the Str\"omgren ($b$+$y$)/2 band \citep{winn07b}.  In \citet{knut09a} we found that the star contributed approximately 1/3 and 1/5th of the total measured increase in flux over half of the planet's orbit in the 8 and 24~\micron~IRAC bands, respectively.

We characterize the behavior of HD~149206 during the epoch of our phase curve observations using the same Str\"omgren ($b$+$y$)/2 band as for HD~189733.  The 564 observations in our analysis were acquired between UT 2005 April 25 and UT 2008 June 7 from an ongoing monitoring program carried out with the T11 0.8 m automatic photometric telescope (APT) at Fairborn Observatory \citep{henry99,eat03}.  Differential brightness measurements were made as the telescope nodded between HD 149026 and three comparison stars of comparable or greater brightness, following the methods described in \citet{henry08}.  These data were used previously by \citet{nutz09}, who computed the periodogram of the time series for periods between $0.5-100$ days and found no evidence for any significant periodicities.  These authors placed an upper limit of 0.001 mag ($0.1\%$) on the peak-to-peak amplitude of any sinusiodal flux variations in this period range.

\begin{figure}
\epsscale{1.2}
\plotone{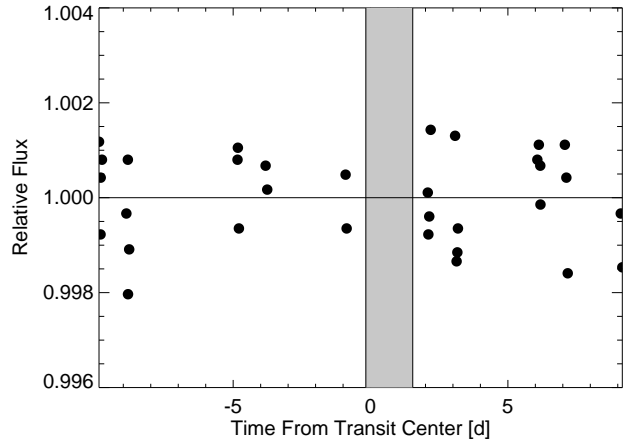} %apt_phot.ps
\caption{Relative flux variation for HD~149026 observed with the T11 0.8~m APT in the Str\o mgren $y$ band (filled circles) around the time of our 8~\micron~\emph{Spitzer} phase curve observations (gray shaded region).  We assume an uncertainty of 0.0009 mag per point ($0.08\%$, calculated as the standard deviation of the fluxes measured in a single night and taking the average of these values for all nights with four or more observations) and fit a linear function of time to these data.  Using this fit, we derive a $2\sigma$ upper limit of $0.005\%$ on the potential increase in the star's flux in Str\o mgren $y$ during our phase curve observations, corresponding to a $2\sigma$ upper limit of $0.0006\%$ at 8~\micron.  This is insignificant relative to the amplitude of the observed phase variation.\label{starspots}}
\end{figure}

The Str\"omgren ($b$+$y$)/2 APT data for HD~149026 indicate that spot-induced variability in HD~149026 is a factor of 30 or more smaller than the variability observed in the same Str\"omgren ($b$+$y$)/2 band for HD~189733 and therefore is unlikely to contribute significantly to the measured 8~\micron~flux increase of $0.0227\% \pm0.0066\%$ for this system.  Because the APT data span the epoch of our 8~\micron~phase curve observations, we derive a more specific constraint on potential flux variations over the 10 days immediately before and after our phase curve observations (see Fig. \ref{starspots}).  We use the Str\"omgren $y$ data alone for this analysis, as they are closer to the infrared wavelengths of our \emph{Spitzer} photometry than the averaged Str\"omgren ($b$+$y$)/2 data.  We set the relative uncertainty on each $y$ band point equal to 0.0009 mag ($0.08\%$, calculated as the standard deviation of the fluxes measured in a single night and taking the average of these values for all nights with four or more observations).  We then fit a linear function of time to the APT $y$ band data and find that the $2\sigma$ upper limit on the increase in HD~149026's flux over the 31 hours of the \emph{Spitzer} observations is $0.005\%$ in Str\"omgren $y$.  Using the conversion factor we derived in \citet{knut08}, we scale the upper limit $y$ band variation to the 8~\micron~band and find a maximum possible 8~\micron~stellar flux increase of $0.0006\%$, a mere $3\%$ of the observed rise in flux in the 8~\micron~band.  As a result we ignore stellar variability for the remainder of this analysis and assume that any changes in the measured 8~\micron~flux are due to the planet and not the star.

\section{Discussion}\label{discussion}

\subsection{Eclipse Fits and Evidence for Variability}

As discussed in the introduction, we compare our transit and secondary eclipse fits to equivalent 8~\micron~observations of the transit of HD 149026 \citep[][obtained in 2007]{nutz09} and the secondary eclipse (H07, obtained in 2005).  We use these older data to search for changes in either the measured radius or the 8~\micron~dayside flux from the planet, and to derive improved estimates for these parameters where possible.

\subsubsection{Transit and Secondary Eclipse Depth Comparisons}\label{depth_comparison}

We find a best-fit planet-star radius ratio of $0.05216\pm0.00078$ when all parameters in our fit are allowed to vary freely (i.e., no assumptions about $a/R_*$ or $i$).  This is within $0.5\sigma$ of the value given by \citet{nutz09}.  Similarly, when we fix $a/R_*$ and $i$ to their best-fit values from \citet{carter09}, we find a planet-star radius ratio of $0.05253\pm0.00076$, $0.6\sigma$ away from the 8~\micron~value given by \citet{carter09}.  We can reduce the uncertainties in our estimate of $R_P/R_*$ by combining the values from our fit and the 8~\micron~transit presented in \citet{nutz09} and re-analyzed by \citet{carter09}; taking the error-weighted average of our value and the updated 8~\micron~value from \citet{carter09}, we find $R_P/R_*=0.05224\pm0.00056$.  Our results are consistent with those of \citet{carter09}, and the smaller error bars provided by our new measurement provides additional support for the trend noted by \citet{carter09}, in which the transit depth between $1-2$~\micron~appears to be marginally deeper than the equivalent values for visible-light photometry \citep{winn08} and the \emph{Spitzer} 8~\micron~bandpass.  \citet{carter09} conclude that this discrepancy is most likely due to systematic errors rather than a real variation in the properties of the planet's atmosphere; the noise in their 1.4~\micron~NICMOS data is a factor of two above the predicted photon noise and its properties are poorly understood. The good agreement between our 8~\micron~transit and that of \citet{nutz09} suggest that the problem most likely lies with the NICMOS data, as the noise in the \emph{Spitzer} data sets is only a factor of 1.28 and 1.15 above the photon noise limit, respectively.

If we compare our new secondary eclipse depth to that reported by H07 in this bandpass, it initially appears as if this quantity might be varying in time.  Our best-fit secondary eclipse depth is $0.0411\% \pm0.0076\%$, a factor of two smaller ($3.0\sigma$) than the value of $0.082\%^{+0.009\%}_{-0.012\%}$ from H07.    Fixing the time of secondary eclipse in our fits to its predicted value results in a slightly shallower eclipse depth, with a net change of $0.006\%$ or $0.8\sigma$ as compared to our original fit.   

It is possible that the removal of large-amplitude detector effects in the older H07 data might have resulted in an over-estimate of the true eclipse depth, in which case our two measurements might in fact be consistent.  Although H07 used the same 0.4~s exposure time and 8~\micron~IRAC subarray as we did, our observations place the star at a single pixel at the center of the subarray whereas the H07 data cycle between nine different nod positions, with a total of twelve complete cycles over 6.4 hours of observations.  Each nod position has a slightly different gain and the move to a new nod position results in a slight reset of the detector ramp effect described in \S\ref{preflash_discussion}, therefore H07 must fit a total of 23 free parameters in their fit as compared to our four free parameters.  They must also remove a ramp spanning the entire data set with a total rise of $1.3\%$, fifteen times larger than their inferred $0.084\%$ decrease in flux during secondary eclipse.

As compared to the H07 fits, our analysis is much simpler.  We do not nod, therefore we have no need to normalize to different nod positions.  We preflash our data to mitigate the ramp and the secondary eclipse occurs at the end of our observations, where the slope is minimal (the best-fit coefficient for the linear function of time in our eclipse fits differs from zero by only $0.2\sigma$ and we obtain the same results when we limit our fits to a simple constant term to describe the out-of-eclipse flux).  We also have more data out of eclipse than H07, which is helpful for determining a robust out-of-eclipse baseline.  Although the uncertainties in our measured eclipse depth are only slightly smaller than those reported by H07 for the 2005 eclipse, we conclude that our new eclipse depth is a more reliable estimate of the planet's 8~\micron~flux for the reasons listed above.

As discussed in \S\ref{harr_data}, we evaluate the robustness of the eclipse depth provided by H07 by downloading the images from 2005 from the \emph{Spitzer} archives and performing our own independent analysis of these data.  Depending on our specific choices for which data to trim and what functional forms to fit, we find eclipse depths ranging from $0.05-0.08\%$, although an eclipse depth around $0.05\%$ is preferred in a majority of our fits (Fig. \ref{harr_plot}).  If we repeat these fits using the same photometry as H07 we find solutions with eclipse depths closer to $0.05\%$ where the $\chi^2$ value is indistinguishable from the best $\chi^2$ value obtained by H07.  Our results suggest that the 2005 data is in fact consistent with our new 2008 observations, and we therefore conclude that there is no evidence for variability in the planet's 8~\micron~flux.   

\begin{figure}
\epsscale{1.2}
\plotone{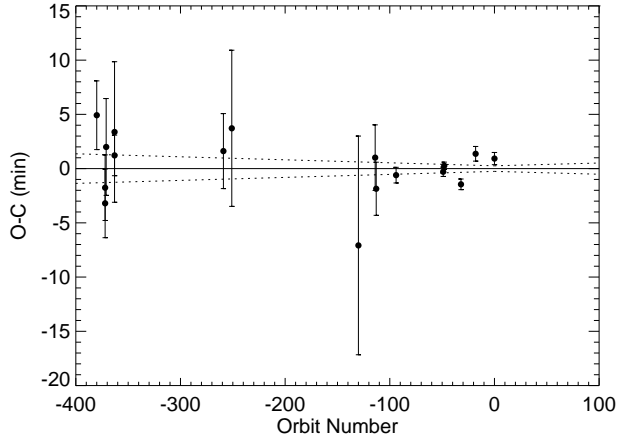} %new_ephemeris_residuals.ps
\caption{Measured transit times for HD 149026b along with our new best-fit ephemeris.  The x axis gives the number of orbits between observations, where our new value is used as the zero point.  The y axis shows the difference between the observed and calculated transit times.  Early points with larger error bars are ground-based data from \citet{sato05}, \citet{charb06}, and \citet{winn08}, while the point at $-94$ orbits is the \emph{Spitzer} transit from \citet{nutz09}, and the four points between $-49$ and $-18$ orbits are \emph{HST} observations from \citet{carter09}.  Dashed lines show the $1\sigma$ uncertainties in the predicted transit times.  The three most recent transits differ from their predicted values by $-3.0\sigma$, $2.0\sigma$, and $1.6\sigma$, respectively, if we neglect the correlated uncertainties in the predicted transit times. \label{o-c}}
\end{figure}

\begin{figure}
\epsscale{1.2}
\plotone{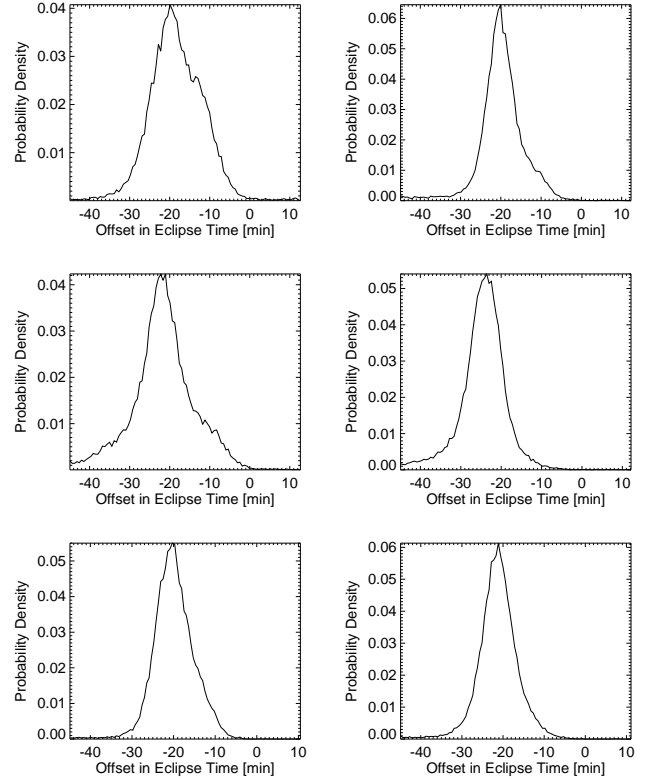} %new_ephemeris_residuals.ps
\caption{Probability distribution of timing offsets from a series of MCMC fits to the secondary eclipse plotted in Fig. \ref{sec_eclipse_plot}.  These fits exclude the first 10 (top left), 11th-20th (top right), 21st-30th (middle left), 31st-40th (middle right), 41st-50th (bottom left), and 51st-60th frames (bottom right) of each set of 64 images.  The differing distributions of timing offsets reflect the presence of correlated noise in the data; the best-fit eclipse depth and corresponding uncertainty, which are less sensitive to correlated noise on shorter time scales, are the same in all of these fits.  We adopt the largest timing errors from this set of fits (middle left histogram, $+7.1$ and $-6.4$ minutes) as a conservative estimate of the uncertainty in the best-fit time of secondary eclipse (see discussion in \S\ref{eclipse_timing}).  For our best-fit eclipse time we keep the original value from our fit to the full data set.\label{timing_hist}}
\end{figure}

\subsubsection{Transit and Secondary Eclipse Timing}\label{eclipse_timing}

Our best-fit transit time occurs $1.18\pm0.62$ minutes earlier than predicted using the ephemeris from \citet{carter09}.  We combine this new transit value with previous transit times from \citet{winn08}, \citet{nutz09}, and \citet{carter09} to derive an improved ephemeris for the planet (Fig. \ref{o-c}).  We find a period of $2.8758925\pm0.0000023$~days and a mid-transit time of $2454597.70645\pm0.00018$, consistent with the ephemeris from \citet{carter09}.  The NICMOS transits described in that work have a higher timing precision than the \emph{Spitzer} transits and the authors note that their measured transit times are marginally inconsistent ($\chi^2=20.16$ with 14 degrees of freedom) with a constant period.  In particular, the two most recent NICMOS transits, which occur 92 and 52 days prior to our observations, differ from their predicted values by $-2.6\sigma$ and $+2.3\sigma$ (1.2 and 1.6 minutes), respectively.  With our new ephemeris we find a $\chi^2$ value of 23.7 with 15 degrees of freedom, with the two most recent NICMOS transits and our new transit differing from their predicted values by $-3.0\sigma$, $2.0\sigma$, and $1.6\sigma$, respectively, where we have neglected the correlated uncertainty in the ephemeris.  

If HD~149026b has a circular orbit, the secondary eclipse should occur half an orbit after the transit, with an additional 45~s delay from the increased light-travel time in the system \citep{loeb05}.  H07 find that their 8~\micron~secondary eclipse occurred $-3.0^{+1.7}_{-2.5}$ minutes from the predicted time (i.e., earlier than predicted).  Our new 8~\micron~secondary eclipse, which was observed three years after that of H07, differs from the predicted time in our nominal fits by $-20.9^{+4.3}_{-4.0}$ minutes with a significance of $4.9\sigma$.  Because transit and secondary eclipse times are particularly sensitive to the presence of correlated noise in the data, we use the ``prayer bead'' method \citep{moutou04} to calculate a second, independent estimate for the uncertainty in our best-fit secondary eclipse time. First we take the residuals from our best-fit secondary eclipse solution and shift them forward by ten points at a time, with the data at the end of the time series wrapping around to the beginning.  After each shift we add the best-fit eclipse function back in and then fit the resulting light curve to determine the best-fit secondary eclipse depth and time.  We repeat this process until the residuals have cycled around to their original positions, and then plot a histogram of the resulting eclipse times.  The resulting distribution is close to Gaussian, consistent with the results of our MCMC analysis, and has a standard deviation of 5.0 minutes.  

As discussed in \S\ref{init_phot}, subarray photometry is recorded in sets of 64 images, and the first $5-10$ frames and 58th frames in each set of 64 tend to have low flux values.  In our initial analysis we concluded that this effect was removed by the background subtraction, but it is possible that some correlated noise remains on these time scales.  As a test we re-ran our secondary eclipse MCMC fits without the first 10 and 58th frames, and found that we obtained identical values for the eclipse depth, but the distribution of values for the best-fit eclipse time was both broader and noticeably asymmetric, with $1\sigma$ uncertainties on the best-fit eclipse time of $^{+7.0}_{-5.7}$.  In order to determine if this change was connected to the presence of the low-background frames, we re-ran the same fits excluding the first 10, 11th-20th, 21st-30th, 31st-40th, 41st-50th, and 51st-60th frames of each set of 64 images (see Fig. \ref{timing_hist}).  In all cases we obtained consistent values for the best-fit eclipse depth and corresponding uncertainties, as well as the best-fit eclipse time, but the uncertainties on this eclipse time ranged from $^{+4.2}_{-4.0}$ (excluding the 51st-60th frames) to $^{+7.2}_{-6.5}$ (excluding the 21st-30th frames).  As a more conservative error estimate, we use the largest of these error bars for the best-fit eclipse time in our subsequent analysis below.  This uncertainty suggests that the measured eclipse time is only marginally inconsistent ($2.9\sigma$ as opposed to our earlier value of $4.9\sigma$) with the predicted value for a circular orbit.  The uncertainty in the secondary eclipse depth is unchanged, as we obtain consistent results in all cases.  We also repeated this same test for the transit fits and found that the uncertainties in the best-fit planet-star radius ratio and transit time were similarly consistent.  We conclude that the only parameter sensitive to this effect is the secondary eclipse time, a fact that is reflected in our new, larger error estimate for this quantity.

Using these new error bars, we find that the offset in our best-fit secondary eclipse time differs from that of H07 by $2.3\sigma$.  In our re-analysis of the H07 data (\S\ref{depth_comparison}) we find a best-fit eclipse time 3.3 minutes later than the value derived by H07, albeit with larger uncertainties ($\pm3.5$ minutes as opposed to H07's value of $^{+1.7}_{-2.5}$ minutes).  Using this value, the timing offset in the 2005 eclipse differs from that of our 2008 eclipse by $2.6\sigma$.  If this offset is confirmed by future observations, it would suggest that the planet's orbit may have changed during the three year interval between the two observations.  

The timing offset observed in our new data corresponds to a value of $-0.0079^{+0.0027}_{-0.0025}$ for $e \cos{(\omega)}$ where $e$ is the planet's orbital eccentricity and $\omega$ is the argument of pericenter.  This is consistent with the radial velocity observations of this system from \citet{sato05} and \citet{wolf07}, which are poorly sampled and thus do not provide strong constraints on the planet's orbital eccentricity.  \citet{madhu09} derive upper limits on $e \sin{(\omega)}$ and $e \cos{(\omega)}$ from a simultaneous fit to the available transit, secondary eclipse, and radial velocity data, but the uncertainty in $e \cos{(\omega)}$ is dominated by the uncertainty in the secondary eclipse time from H07 and their value for $e \sin{(\omega)}=0.109^{+0.042}_{-0.068}$ is still poorly constrained relative to our estimate for $e \cos{(\omega)}$.   If real, the observed change in $e \cos{(\omega)}$ could be due to either a changing orbital eccentricity or a precessing orbit (i.e., a change in $e$ or in $\omega$).  The addition of four new secondary eclipses observed as part of an ongoing \emph{Spitzer} target of opportunity program (J. Harrington 2009, private communication) should either confirm or disprove the presence of the timing offset suggest by our data, as well as constraining any temporal variability in this quantity.    

\begin{figure}
\epsscale{1.2}
\plotone{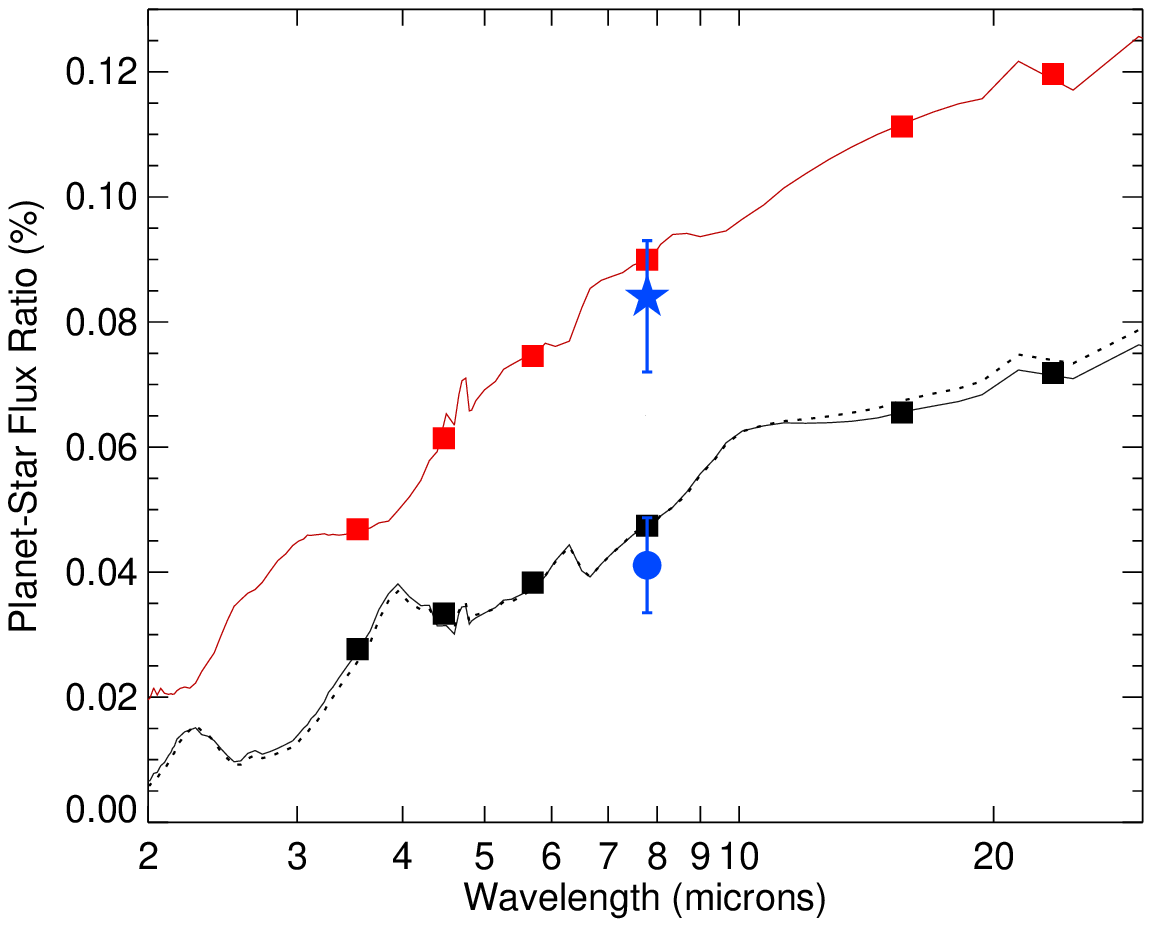} %final_day_spectrum.ps
\caption{Best-fit 8~\micron~secondary eclipse depths from this paper (filled circle) and from \citet{harr07} (filled star) with $\pm1\sigma$ error bars overplotted.  We re-analyze the \citet{harr07} data in this paper and find a smaller value consistent with our new secondary eclipse depth.  Two dayside atmosphere models for HD~149026b are shown for comparison (these are similar to the models in Fortney et al. 2006, but these models have been updated to match the revised system parameters from Carter et al. 2009). The black model (solid line) assumes half the incident flux is transported to the night side (i.e. no day-night temperature gradient), does not include TiO opacity and therefore has no inversion, and shows water and CO bands in absorption.  The other black model (dotted line) makes the same assumptions but uses a $30\times$ solar metallicity atmosphere for the planet instead of solar metallicity.  This assumption has a minimal effect on the resulting planet-star flux ratio.  In contrast to the two black models, the red model absorbs and re-radiates all of the incident flux from its star on the day side (this assumes effectively zero flux from the cold night side), includes TiO opacity, and has formed a temperature inversion with water and CO features in emission.  The filled squares give the predicted planet-star flux ratios integrated over the six Spitzer bandpasses (3.6, 4.5, 5.8, 8.0, 16, and 24~\micron) for each of these models.\label{spectrum}}
\end{figure}

\subsection{Comparisons to 1D Atmosphere Models}\label{1Dmodels}

\emph{Spitzer} observations of secondary eclipses for close-in hot Jupiters \citep[e.g.][]{dem05,dem06,charb05,charb08,harr07,rich07,grillmair07,grillmair09,mach08,knut08,knut09b} suggest that their atmospheres can be divided into two classes, depending on whether or not they have a temperature inversion and water emission (as opposed to water absorption) features in the near- to mid-IR.  Planets with inversions appear to require an additional absorber at altitude, preferably with a large cross-section in visible light where the stars emit most of their flux, in order to create the inversion.  Several authors \citep{hub03,fort06,fort08,burr07b,burr08} have suggested that gas-phase TiO might provide the required opacity, although cold trap effects would reduce the amount of TiO in the upper atmosphere to much lower levels unless there is fairly vigorous mixing in this region \citep{spiegel09,show09}.  In this picture planets with sufficiently high levels of incident flux would have gas-phase TiO and would form inversions, whereas cooler planets would not.  Although this theory works well for two of the three hot Jupiters with confirmed inversions \citep{knut08,knut09b}, it breaks down for XO-1b \citep{mach08}, which is too cool for gas-phase TiO and yet appears to have an inversion.   \citet{zahnle09} recently proposed an alternative theory where photochemistry in the upper atmosphere might lead to the production of enough disequilibrium sulfur compounds to form an inversion.  In this case the authors find that the formation rates for the sulfur compounds of interest appear to be independent of stellar insolation, atmospheric metallicity, and temperature in the $1200-2000$~K range spanned by their models.  Although these models correctly predict that XO-1b should have an inversion, they fail to explain why HD~189733b and TrES-1 do not.

HD~149026b is a clear candidate for an inversion in either scenario, as it receives enough incident flux in the zero-albedo case to have gas-phase TiO in the hotter regions of its dayside atmosphere, and the sulfur photochemistry model suggests that all close-in planets should have inversions.  Although this planet has an equilibrium temperature of $1740$~K, the 8~\micron~secondary eclipse depth measured by H07 corresponds to an 8~\micron~brightness temperature of $2300\pm200$~K.  In order to match this high dayside flux, models must keep as much energy as possible on the day side (i.e., a large day-night temperature gradient).  Models with a temperature inversion and water in emission also provide a better match to the data \citep{fort06,burr08}, as they tend to have higher 8~\micron~fluxes than their non-inverted counterparts, which have water absorption features in this wavelength range (see Fig. \ref{spectrum}). With only one wavelength of data, however, we cannot directly verify the presence or absence of a temperature inversion in HD 149026b's dayside atmosphere.

We measure a considerably smaller 8~\micron~secondary eclipse depth in our new 8~\micron~observations; our value as plotted in Fig. \ref{spectrum} is half that reported by H07, with a brightness temperature of only $1440\pm150$~K.  This value is best matched by models with strong circulation between the two hemispheres (i.e., half the incident flux is transported to the night side), as this serves to reduce the predicted dayside temperatures.  Models with absorption features in the 8~\micron~bandpass, such as the non-inverted model in Fig. \ref{spectrum}, also tend to provide a better fit to the data, but the observed 8~\micron~flux could also be matched by a model with blackbody emission and a non-zero albedo.  It is risky to extrapolate too much on the basis of observations at a single wavelength, and definite confirmation of the presence or absence of an inversion must await the addition of observations at other wavelengths.  We will carry out full-orbit phase curve observations of this system in the 3.6 and 4.5~\micron~IRAC bands, which have proven to be particularly useful for diagnosing the presence of such temperature inversions \citep[e.g.][]{knut08}, as part of \emph{Spitzer's} extended warm mission from $2009-2011$.

Although our reanalysis of the H07 data is consistent with a constant 8~\micron~secondary eclipse depth from $2005-2008$, it is interesting to speculate on the possibility that HD 149026b is time variable.  Very few planets have more than one secondary eclipse observation in the same IRAC channel, but this planet has been observed four times at 8~\micron~and twice at 5.8~\micron~(two of the 8~\micron~secondary eclipses and two new 5.8~\micron~secondary eclipses were observed as part of an ongoing \emph{Spitzer} target of opportunity program and are currently unpublished; J. Harrington 2009, private communication).  These new data, when combined with our eclipse and that of H07, will allow for a search for variability on a time scale of four years with multiple epochs of observations, and will be sensitive to smaller variations than those detectible in the current data.  

Time variability is common in planetary atmospheres, and several hot Jupiter models have suggested highly time-variable flows \citep[e.g.,][]{cho03,cho08,lang08,menou09}.  Obtaining large-amplitude time variability on the global scale requires the existence of dynamical instabilities that (i) have length scales comparable to (or larger than) a planetary radius, and (ii) having instability growth times shorter than the timescales of any processes that might damp the instabilities; in practice this probably requires growth times shorter than the relevant radiative time constants.  To date, most theoretical studies suggesting that global-scale variability could be significant adopt two-dimensional models; some of these models essentially ensure a time-variable flow by adopting a turbulent initial condition \citep{cho03,cho08,rau08} while others generate highly fluctuating flows even when initialized from rest \citep{lang08}.  On the other hand, most three-dimensional models of synchronously rotating hot Jupiters on circular orbits published to date produce relatively steady flow patterns that do not exhibit large-amplitude global-scale variability \citep{show02,coop05,coop06,show08,show09,dobb08}.  One three-dimensional study produced highly time variable flows \citep{menou09}; the different behavior might result from the very different forcing scheme adopted in Menou \& Rauscher (and their placement of the bottom surface at relatively shallow pressure) as compared to previous 3D studies.  These theoretical predictions for variability will likely evolve as models explore a wider parameter space in forcing schemes and planetary parameters, while the availability of multiple epochs of secondary eclipse observations for a few select planets should provide the first observational tests for such variability.  In \citet{agol08} we present one such search for variability in the 8~\micron~emission from HD~189733b, where we conclude that this planet's emission appears to be constant at the level of $10\%$ over the five secondary eclipses observed.

\subsection{The 3D Picture: A Global Energy Budget}\label{3Dmodels}

Rather than extrapolating the nightside temperature based on the 1D dayside models plotted in Figure \ref{spectrum}, we can use our measurement of the planet's phase variation to directly characterize this quantity.  We measure an increase in flux of $0.0227\% \pm0.0066\%$ over the course of our observations; when combined with our new value for the secondary eclipse depth, which gives the dayside flux, this corresponds to a minimum hemisphere-averaged flux that is $45\% \pm19\%$ of the maximum hemisphere-averaged flux.  If we assume the planet emits as a blackbody and solve for the temperature needed to match the maximum and minimum hemisphere-averaged planet-star flux ratios (Table \ref{temp_table}), this corresponds to a maximum (dayside) hemisphere-averaged brightness temperature of $1440\pm150$~K and a minimum (nightside) hemisphere-averaged brightness temperature of $960\pm240$~K.  The day-night temperature gradient in this case is then $480\pm140$~K, where the uncertainty in this quantity simply reflects the uncertainty in the measured rise in flux during these observations.  The uncertainty in the dayside temperature is derived from the uncertainty in the depth of the secondary eclipse, while the uncertainty in the night-side temperature is a function of both the uncertainty in the secondary eclipse depth and the day-night flux difference and is therefore larger than either of these quantities.  The resulting day-night temperature gradient for HD~149026b is larger than the approximately $200\pm50$~K gradient observed for HD~189733b \citep{knut09a}, but this difference is not statistically significant.  

\begin{figure}
\epsscale{1.2}
\plotone{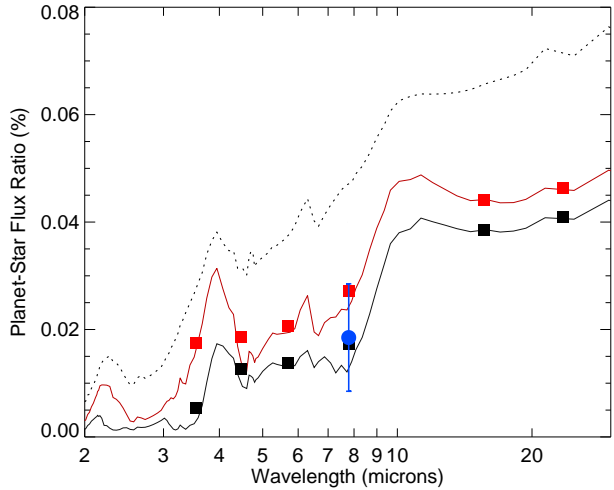} %final_night_spectrum.ps
\caption{This plot shows the best-fit 8~\micron~night-side flux from our phase curve fit with $\pm1\sigma$ error bars overplotted.  The two overplotted models show the predicted emission from an object in isolation (i.e., zero incident flux) with an effective temperature of 1300~K (black line) or 1600~K (red line).  The dotted black line is the best-fit dayside model from Fig. \ref{spectrum}.  The filled squares give the predicted planet-star flux ratios integrated over the six Spitzer bandpasses (3.6, 4.5, 5.8, 8.0, 16, and 24~\micron) for each of these models.\label{night_spectrum}}
\end{figure}

If we instead use the two-hemisphere model fit to the phase curve from \S\ref{phase_var_fits} to determine the maximum and minimum hemisphere-averaged fluxes, we find a larger gradient of $780\pm240$~K between these two hemispheres.  The maximum flux in this model occurs after the end of our observations (see Fig. \ref{phase_curve_plot}), and the increased temperature gradient is a direct result of this extrapolated maximum flux value.  We prefer the linear fit as a more conservative estimate of the day-night temperature gradient, although it is in some sense a lower limit as the planet still appeared to be increasing in brightness at the end of our observations.

Although both the minimum and maximum hemisphere-averaged temperatures are lower than HD~149026b's $1770$~K equilibrium temperature assuming an albedo of zero and uniform ($4\pi$) re-radiation of incident flux, this does not necessarily mean that there is a problem with its energy budget.  If the planet has an emission spectrum with water absorption it will appear correspondingly fainter in our 8~\micron~bandpass, and in fact we see that the 1D model spectrum in Fig. \ref{spectrum} is a good match for our dayside planet-star flux ratio.  This model has a Bond albedo of near zero and an effective temperature of $1750$~K, and it assumes that half the incident flux is redistributed to the night side.  We obtain comparable fits with both solar-metallicity and 30x solar metallicity models for the planet's atmosphere.  

For the planet's night side, which receives no incident flux from the star, we use a model for an object in isolation and vary the effective temperature in order to match the observed 8~\micron~flux (see Fig. \ref{night_spectrum}).  We find that a model with $T_{eff}=1300$~K provides the best fit, although $T_{eff}=1600$~K is still within the $1\sigma$ uncertainty for the night-side flux.  Using an effective temperature of 1750~K for the day side and 1300~K for the night side we find that the planet emits $60\%$ as much flux as it receives from its parent star in the zero-albedo case.  In order to balance these two quantities the planet must then have a Bond albedo of $40\%$.  If we carry out the same calculation with a $1600$~K night side we find a Bond albedo of $17\%$.  This is not a fully self-consistent solution, as we do not re-run our dayside model with a non-zero albedo (which could perhaps come from reflecting clouds) and the appropriate redistribution required to match the nightside flux, but the difference is insignificant in light of the current uncertainties in the measured day- and night-side fluxes and the limited wavelength range of the available data.  

Upper limits on reflected light from hot Jupiters such as HD~209458b \citep{rowe08} suggest that their albedos are small (the $1\sigma$ upper limit for the geometric albedo of HD~209458b is $0.08$), a conclusion that is in good agreement with the predictions of atmosphere models for these planets \citep[e.g.][]{burr08b}.  CoRoT-1b was also recently found to have a geometric albedo $<0.20$ \citep[this planet still has some thermal emission in the red CoRoT band of these observations, and the observed secondary eclipse signal could be due to either reflected light or thermal emission from the planet;][]{snellen09}, placing it in a regime similar to that of HD~209458b.  HD~149026b appears to be an exception to this trend, although it is worth remembering that the planet emits most of its flux at shorter wavelengths where it may be brighter than predicted by our atmosphere models.  If this was the case, the planet's emission could still be consistent with an albedo of zero.  Additional observations at shorter wavelengths should allow us to test the validity of our atmosphere models and to provide tighter constraints on the planet's Bond albedo.  

\section{Conclusions}\label{conclusions}

In this work we present a revised value for the 8~\micron~dayside flux from the hot Saturn HD~149026b and the first estimate of the nightside flux and corresponding day-night temperature gradient.  Our new secondary eclipse depth is half that reported by H07 in the same 8~\micron~bandpass for observations obtained in 2005, and is best matched by atmosphere models with strong day-night circulation, water absorption in the 8~\micron~bandpass, a non-zero albedo, or some combination of the three.  Our conclusion stands in direct contrast to that reached by H07, who concluded that the planet's apparent high flux in the 8~\micron~bandpass was best described by a model with a large day-night temperature gradient and/or water emission features.  Our re-analysis of the H07 data indicates that the 2005 data are in fact consistent with our new, smaller 8~\micron~eclipse depth, and we conclude that there is no evidence for variability in the planet's 8~\micron~flux over the three years between these two observations.  There have been two additional 8~\micron~secondary eclipse observations (currently unpublished) obtained within the past year; these data should provide a more rigorous test for any potential variability in HD~149026b's dayside emission.  Observations of the secondary eclipse at 3.6 and 4.5~\micron, which have been scheduled as part of \emph{Spitzer's} extended two-year warm mission, will provide an important complement to these data and should allow us to confirm the presence or absence of a temperature inversion.

Although we find no evidence for changes in the planet's dayside flux, our new secondary eclipse occurs $-20.9^{+7.2}_{-6.5}$ minutes earlier than the predicted time, $2.3\sigma$~earlier than the 2005 eclipse from H07 (the best-fit eclipse time from our re-analysis of the 2005 data is comparable to that reported by H07).  The timing offset for our new secondary eclipse corresponds to a value of $-0.0079^{+0.0027}_{-0.0025}$ for $ecos(\omega)$ where $e$ is the planet's orbital eccentricity and $\omega$ is the argument of pericenter.  By mid-2011 \emph{Spitzer} will have observed eight secondary eclipses for this planet (one each at 3.6 and 4.5~\micron, two at 5.8~\micron, and four at 8.0~\micron); these data will allow us to confirm or disprove the presence of an offset in the secondary eclipse times and to search for any evidence of temporal variability.

Our measurement of the planet's phase curve indicates that the minimum hemisphere-averaged 8~\micron~flux for this planet is $45\% \pm19\%$ of the maximum hemisphere-averaged flux.  We use our estimates of the day- and night-side fluxes to construct an energy budget for the planet and conclude that it appears to emit less flux than it absorbs from its parent star, corresponding to a non-zero Bond albedo, although the models are poorly constrained by the current data.  Because we only have measurements of the day- and night-side fluxes in a single bandpass located on the long-wavelength tail of the planet's emission spectrum, our observations may still be consistent with a zero albedo if the planet is brighter than predicted at shorter wavelengths.  Although our models have similar predictions for solar and $30\times$ solar atmospheric metallicities, variations in the abundances of specific elements might still affect our predictions for the 8~\micron~bandpass.  

Over the next two years we will obtain full-orbit phase curve observations of HD~149026b at 3.6 and 4.5~\micron~as part of \emph{Spitzer's} extended warm mission; these wavelengths are close to the peak of the planet's emission spectrum and will therefore provide much tighter constraints on its energy budget.  The full phase coverage will also allow us to better resolve the locations of hot and cold regions in the planet's atmosphere. These observations will take place as part of a larger program to provide multi-wavelength characterizations of the phase curves of six extrasolar planets, including HD 189733b, HD 209458b, HD 149026b, HAT-P-2, HAT-P-7, and GJ 436b, and comparisons between these systems will allow us to investigate the importance of irradiation, rotation rate, surface gravity, eccentricity, and stellar metallicity in determining the pressure-temperature profiles and dynamic meteorology of these unusual atmospheres.

\acknowledgements

This work is based on observations made with the \emph{Spitzer Space Telescope}, which is operated by the Jet Propulsion Laboratory, California Institute of Technology, under contract to NASA.  Support for this work was provided by NASA through an award issued by JPL/Caltech.

\end{document}